\documentclass[twocolumn,showpacs,preprintnumbers,amsmath,amssymb]{revtex4}

\usepackage{epsfig}
\usepackage{graphicx}
\usepackage{dcolumn}
\usepackage{bm}

\begin{document}

\title{Effective temperature for finite systems}

\author{Steven Huntsman}%
\email{sh@eqnets.com}
\address{%
Equilibrium Networks\\
Alexandria, Virginia 22314
}%
\altaddress{%
Also at the Department of Physics, Naval Postgraduate School\\
Monterey, California 93943
}%

\date{\today}

\begin{abstract}
Under the \emph{Ansatz} that the occupation times of a system with finitely many states are given by the Gibbs distribution, an effective temperature is uniquely determined (up to a choice of scale), and may be computed \emph{de novo}, without any reference to a Hamiltonian for empirically accessible systems. As an example, the calculation of the effective temperature for a classical Bose gas is outlined and applied to the analysis of computer network traffic. 
\end{abstract}

\pacs{05.70.-a, 65.40.Gr, 07.05.Kf, 89.70.+c}
\maketitle

\paragraph{Introduction.}

It is commonplace to characterize communications channels and chaotic and complex systems through the use of some type of entropy measure. However, further dynamical characterizations are often discipline-specific: e.g., fidelity for quantum error-correcting codes, rates of divergence or Lyapunov exponents, attractor reconstruction methods, etc. \cite{BS} While system-specific differences are to be expected, it is always desirable to have more versatile tools for characterization. This viewpoint suggests the potential to apply fundamental concepts from statistical physics beyond entropy that have yet to be developed in subject areas outside of their historically derived roles. Meanwhile, a common feature in the methods listed above (and indeed the fundamental property of any dynamical characterization, regardless of its actual purpose) is a functional dependence on the dynamical time evolution. The emergence of a general dynamical measure would complement the widely adopted class of characterizations constructed from density matrices and their ilk, of which entropies are the most prominent. An ideal candidate for such a measure is an effective temperature \cite{CJ}. We review the construction of such a temperature here. Its form (see equation \eqref{eq:temperature}) is uniquely determined by the requirement of consistency with equilibrium statistical physics, and its applicability is enhanced by the virtue that it is a function only of the state occupation times (equivalently, the state probabilities and recurrence time). In consequence, we arrive at the following result of interest: rather than playing the role of an environmental parameter in calculations with detailed models, temperature can be regarded as an intrinsic quantity that can be computed \emph{de novo} purely on the basis of time averages. By closing the Gibbs relation, the same point of view can be extended to the state energies. We expect that the framework presented here can be fruitfully applied to the analysis of generic empirically accessible systems.

Our work is a natural outgrowth of several recent developments, one set of which is reviewed in \cite{CJ}. It is now well known how to calculate the temperature from time averages for a given Hamiltonian \cite{R}, \cite{Bannur}, \cite{JAE}. In a more abstract setting, it has also been illustrated how energies and probabilities or time averages suffice \cite{GMM}, \cite{Carati}. 

The requirement of empirical accessibility for our framework appears to exclude much of the traditional domain of statistical physics from applications, but it also suggests completely new applications in the spirit of information theory, e.g. to data from networks and markets, or more generically to large data sets. One such application of immediate interest is to computer network traffic: the use of thermal variables other than entropy to describe computer network traffic has been considered previously \cite{Burgess}, \cite{FH}. Another area worthy of investigation is temperature in granular media \cite{MJ}, \cite{CBL}.

\paragraph{Two sets of coordinates.}

Let us make it clear at the outset that our goal is to describe systems using the idiom of equilibrium statistical physics in the canonical ensemble. Consider a finite system with $n$ states, with nonzero occupation probabilities $\{p_k\}_{k=1}^{n}$ and characteristic recurrence time $t_{\infty}$. We begin in earnest by noting that we may stipulate that the (otherwise as yet undetermined) state energies satisfy the zero-point constraint
\begin{equation}
\label{eq:E0}
E_1 + \cdots + E_n = 0.
\end{equation}
(Note that this is not the same as fixing the internal energy.) Keeping \eqref{eq:E0} in mind, we introduce an empirical temperature parameter $\Theta \equiv k_BT$ and two sets of coordinates. The \emph{experimentalist's coordinates} are given by the characteristic state occupation times:
\begin{equation}
\bm{t} := (t_1,\dots,t_n),
\end{equation}
with $t_k := t_{\infty}p_k$ (or in shorthand $\bm{t} = t_{\infty}\bm{p}$), and the \emph{theorist's coordinates} are given by
\begin{equation}
\bm{H} := (E_1,\dots,E_n,\Theta).
\end{equation}
$\bm{t}$-coordinates lie in $X_{\bm{t}} := \{\bm{t} : t_k > 0, \forall k\}$; $\bm{H}$-coordinates typically (but not necessarily always \cite{Ramsey}) lie in $X_{\bm{H}} := \{\bm{H} :  (\Theta > 0) \land (E_1 + \cdots + E_n = 0)\}$.
	
It is natural to ask how to transform from one set of coordinates to the other. The transformation from $X_{\bm{H}}$ to $X_{\bm{t}}$ is the usual aim of statistical physics. The main issue in this direction is the computation of the partition function, which is trivial for finite systems in the sense that the computation can be effected in $O(n)$ arithmetic steps. The other direction, from $X_{\bm{t}}$ to $X_{\bm{H}}$, is far less explored: navigating it raises the prospect of bringing the tools of thermodynamics and statistical physics to bear on a host of real-world problems that may have hitherto appeared to lie outside their domain. 

Without any constraints (e.g., without fixing the internal energy of the system \emph{\`{a} la} Jaynes \cite{J}), the Gibbs {\sl Ansatz} for the state occupation probabilities is underdetermined, i.e., the assignment (with $\beta := \Theta^{-1}$)
\begin{equation}
\label{eq:Gibbs}
t_k/t_{\infty} =: p_k^{(\bm{t})} \equiv p_k^{(\bm{H})} := e^{-\beta E_k}/Z,
\end{equation}
still does not uniquely specify a point in $X_{\bm{H}}$, but only a ray, since $c t_k / c t_{\infty} = t_k/t_{\infty}$ and $(\beta/c)(c E_k) = \beta E_k$.

\paragraph{Preliminaries.}

To begin a condensation and refinement of \cite{F1}, the equations \eqref{eq:E0} and \eqref{eq:Gibbs} imply that
\begin{equation}
\label{eq:betaEk}
\beta E_k = \frac{1}{n}\sum_{j=1}^{n} \log \frac{p_j}{p_k},
\end{equation}
and we can use this to obtain $\beta \bm{H} = (\beta E_1,\dots,\beta E_n,1)$. The angle $\phi$ between the unit vector $e_{\Theta} := (0,\dots,0,1)^*$ and $\bm{H}$ is given by
\begin{equation}
\cos \phi := \left \langle e_{\Theta}^*, \frac{\beta \bm{H}}{\lVert \beta \bm{H} \rVert} \right \rangle = \frac{1}{\lVert \beta \bm{H} \rVert}.
\end{equation}

The temperature is therefore given by
\begin{equation}
\label{eq:normhcosphi}
\Theta(\bm{t}) = \lVert \bm{H} \rVert \cos \phi,
\end{equation}
where
\begin{equation}
\label{eq:cosphi}
\cos \phi = \frac{1}{\lVert \beta \bm{H} \rVert} = \left ( \sum_{k=1}^{n} \left ( \beta E_k \right )^{2} + 1 \right )^{-1/2}.
\end{equation}
To obtain $\Theta(\bm{t})$, it remains only to compute $\lVert \bm{H} \rVert$. To do this we shall require two preliminary results. The first result is a scaling behavior of the type suggested by the Wick rotation/correspondence $\beta \leftrightsquigarrow it$ in finite temperature field theory: this will be examined from several points of view. The second result is a geometrical result called the \emph{radial foliation lemma}.

\paragraph{POV 1: dimensional analysis.}

It is clear from dimensional considerations that $T \equiv \Theta/k_B$ must depend on some governing parameter $x$ besides $k_B$ and $\bm{t}$. W/l/o/g, $x$ carries units of action and may be set to $\hbar$, so $T \equiv f(\hbar, k_B, \bm{t})$ for some $f$. Now the Buckingham $\Pi$-theorem \cite{Buckingham}, \cite{Barenblatt} implies that there is a non-dimensional function $\Phi(\bm{p})$ such that $T = \hbar k_B^{-1} t_\infty^{-1} \Phi(\bm{p})$, i.e.
\begin{equation}
\label{eq:dimensional scaling}
\Theta = \hbar t_\infty^{-1} \Phi(\bm{p}).
\end{equation}
Therefore $\Theta$ scales as $1/t_\infty$. Moreover, it is clear from \eqref{eq:dimensional scaling} that using a unit of action other than $\hbar$ as a governing parameter merely amounts to a change of scale for $\Theta$. 

\paragraph{POV 2: two ideal gas systems.}

Consider the following thought experiment, in which we have two systems, comprised respectively of finite ideal gas samples with particle masses $m$ and $Cm$, each in identical freefalling containers in contact with its own isotropic thermal bath, and with the same initial conditions in phase space. Insofar as the system microstates are not of interest in equilibrium, the systems may be respectively described by, e.g. the quintuples $(m, v, t_\infty; p, \beta)$ and $(Cm, v/C, Ct_\infty; p, C\beta)$, where the rms velocities and momenta are indicated. Each system follows the same trajectory through phase space, albeit at rates that differ by constant factors, and we see that $\Theta$ scales as $1/t_\infty$ for ideal gases, and hence (by means of a suitable coupling with an ideal gas bath) for general systems also.

\paragraph{POV 3: extended canonical transformation.}

Consider a Hamiltonian $H(X,P)$. Dilating the dynamical rate by a factor $C$ has the effect that $t_{\infty} \mapsto t'_{\infty} = t_{\infty}/C$ and amounts to a change of units, as it also induces the extended canonical transformation $X \mapsto X' = X, P \mapsto P' = CP, H \mapsto H' = CH$. Meanwhile, $\beta \mapsto \beta'$ and $e^{-\beta H}$ is invariant, so it follows that $\beta' = \beta/C$ and $\Theta$ scales as $1/t_\infty$. 

\paragraph{POV 4: thermal time hypothesis.}

Let $H$ be a Hamiltonian on a finite-dimensional Hilbert space. The thermal density matrix is $\omega = Z^{-1}\mbox{Tr}(e^{-\beta H})$, and the time evolution of an observable $A$ is given as usual by $\tau_t(A) = e^{iHt/\hbar} A e^{-iHt/\hbar}$. 

Now the one-parameter \emph{modular group} of $\omega$ that appears in the Tomita-Takesaki theory of von Neumann algebras \cite{BR} can be shown to coincide with the time evolution group \cite{CR}: if $s$ is the modular parameter and $t$ is the physical time, then 
\begin{equation}
\label{eq:TTH}
t = \beta s/\hbar.
\end{equation}
In particular, $s$ does not depend on $\beta$.

The \emph{thermal time hypothesis} (TTH) articulated by Connes and Rovelli \cite{CR} (see also \cite{MR}, \cite{Martinetti}, \cite{Rovelli2}, \cite{T}) states that physical time is determined by the modular group, which is in turn determined by the state. The TTH simultaneously inverts and generalizes the Kubo-Martin-Schwinger condition \cite{GV}, so that temperature provides the physical link between time evolution and equilibria. Moreover, the TTH implies Hamiltonian mechanics and the Gibbs postulate. Its key implication here though is simply \eqref{eq:TTH}, which implies that $\Theta$ scales as $1/t_\infty$.

\paragraph{The radial foliation lemma.}

We now establish the radial foliation lemma. As mentioned after \eqref{eq:Gibbs}, the state probabilities are constant on rays in both $X_{\bm{t}}$ and $X_{\bm{H}}$. Let $e_r^{(\bm{t})}, e_r^{(\bm{H})}$ denote radial unit vectors in $X_{\bm{t}}$ and $X_{\bm{H}}$, respectively. We will make the mild assumption that any relevant partial derivatives exist on the interiors of $X_{\bm{t}}$ and $X_{\bm{H}}$. It is easy to see that 
\begin{equation}
d\bm{t} = d\lVert \bm{t} \rVert e_r^{(\bm{t})} \iff d\bm{H} = d\lVert \bm{H} \rVert e_r^{(\bm{H})}, \nonumber
\end{equation}
i.e., one of the differentials is purely radial iff both are, and in such an event the probabilities remain constant; likewise, one of the differentials $d\bm{t}$, $d\bm{H}$ is purely angular iff both are, and in such an event the probabilities change. We therefore obtain the lemma: that any reasonably smooth map between $X_{\bm{t}}$ and $X_{\bm{H}}$ that respects \eqref{eq:Gibbs} sends rays and sphere orthants in $X_{\bm{t}}$ to rays and hemispheres in $X_{\bm{H}}$, respectively.

\paragraph{The effective temperature.}

By the radial foliation lemma, $\left \lVert \bm{H}(\bm{t}) \right \rVert = \lVert \bm{H}(\lVert \bm{t} \rVert \hat{1}) \rVert \equiv \Theta \left ( \lVert \bm{t} \rVert \hat{1} \right)$, so that \eqref{eq:normhcosphi} takes the form
\begin{equation}
\label{eq:UT}
\Theta(\bm{t}) = \Theta \left ( \lVert \bm{t} \rVert \hat{1} \right) \cos \phi,
\end{equation}
where $\hat{1}$ is the unit vector with all components equal to $n^{-1/2}$. Meanwhile, the form implied by the scaling behavior of $\Theta$ w/r/t $t_\infty$ for the term $\Theta \left ( \lVert \bm{t} \rVert \hat{1} \right)$ on the RHS of \eqref{eq:UT} is 
\begin{equation}
\label{eq:KMSUT}
\Theta \left ( \lVert \bm{t} \rVert \hat{1} \right) = K/\lVert \bm{t} \rVert,
\end{equation}
where $K$ is a fixed constant with dimensions of action (say, $\hbar$) whose value we may set to unity w/l/o/g.

In the language of an earlier paper \cite{FH}, \eqref{eq:KMSUT} is a \emph{topologically admissible uniform temperature map}. This means first that the LHS of \eqref{eq:KMSUT} is bijective in $\lVert \bm{t} \rVert$, which is essentially the zeroth law of thermodynamics, and second that $\Theta \rightarrow 0$ (resp., $\infty$) as $t_{\infty} \rightarrow \infty$ (resp., $0$). Combined with bijectivity, this implies in particular that the LHS of \eqref{eq:KMSUT} must be monotone decreasing in $\lVert \bm{t} \rVert$.

Combining \eqref{eq:cosphi} and \eqref{eq:KMSUT} using \eqref{eq:UT} leads to a simple formula for $\Theta$:
\begin{equation}
\label{eq:temperature}
\Theta(\bm{t}) = \frac{K}{t_{\infty} \lVert \bm{p} \rVert} 
\left ( \sum_{k=1}^{n} \left ( \frac{1}{n} \sum_{j=1}^{n} \log \frac{p_j}{p_k} \right )^{2} + 1 \right )^{-1/2}
\end{equation}

\begin{figure}[htbp]
\includegraphics[width=80mm,keepaspectratio]{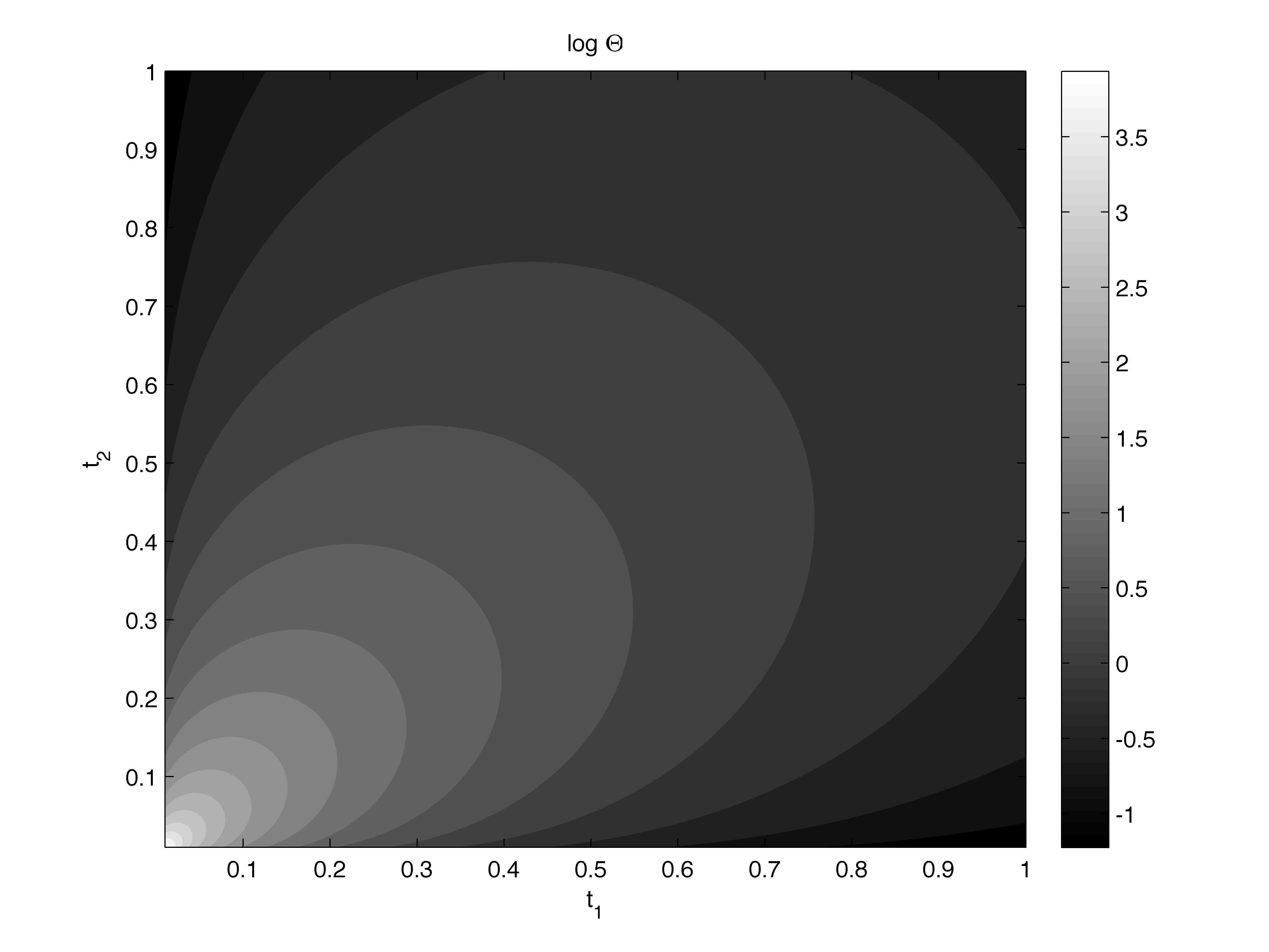}
\caption{ \label{fig:semi} Contour plot of (the logarithm of) $\Theta$ for a two-state system as a function of the occupation times, with $t_{\infty} = t_1 + t_2$. The picture in higher dimensions is qualitatively similar.}
\end{figure}

\paragraph{Temperature of a classical Bose gas.}

We now consider the example of a classical Bose gas \cite{Balantekin}. (The equivalent system in queueing theory is known as a Gordon-Newell or closed Jackson queue \cite{GN}, \cite{Jackson}, and similar but distinct model physical systems include generalized Ehrenfest models \cite{Ehrenfest}, \cite{Siegert}.) The gas consists of $b$ particles with $B$ internal states in continuous time, where the particles independently execute state transitions. A microstate of the gas (versus a state of a single particle) is given by an element of the space $X \equiv X_{B,b} := \{\alpha \in \mathbb{Z}^B : (\lvert \alpha \rvert = b) \land (\alpha > 0)\}$, where we employ standard notation for multi-indices. 

Write $q_j$ for the transition rate from the $j$th (internal particle) state, and write $R_{jk}$ for the probability of a particle transition to the $k$th state from the $j$th state. For $j \neq k$, the probability that a particle attempts to transition from state $j$ to state $k$ (regardless of whether or not any particles are actually in state $j$) is then given in the limit by
\begin{equation}
\label{eq:generator}
Q_{\alpha, \alpha - e_j + e_k} \overset{j \ne k}{:=} \lim_{\Delta t \rightarrow 0}\frac{\mathbf{P}\left( \alpha \overset{\Delta t}{\longrightarrow} \alpha - e_j + e_k \right)}{\Delta t} = q_jR_{jk}.
\end{equation}

Now $R$ is a stochastic matrix and (we shall assume it to be irreducible, so that) it possesses a unique invariant distribution satisfying $\pi = \pi R$. Set $\eta_j := \pi_j/q_j$. It is not hard to show (see appendix) that the induced invariant distribution on $X_{B,b}$ is of the form
\begin{equation}
\mathbf{P}(\alpha) = \eta^{\alpha}/z(\eta). 
\end{equation}
$z$ is a so-called complete homogeneous symmetric function. It can be shown (see e.g. exercise 7.4 of \cite{S}, although this result is apparently not well known in either the physics or queueing theory literature) that
\begin{equation}
\label{eq:CHSF}
z(\eta) = \sum_{k=1}^{B}\frac{\eta_k^{B+b-1}}{\prod_{m \neq k}(\eta_k - \eta_m)}
\end{equation}
if the components of $\eta$ are distinct. By omitting duplicate components and reducing $B$ accordingly, the appropriate generalization of \eqref{eq:CHSF} follows.

Write $\partial(J) := \{\alpha: \alpha_j = 0 \iff j \in J\}$ and $J_{(\#J)} \equiv J$. Now the recurrence time is given by the average inverse net transition rate, viz. $t_{\infty} = \sum_\alpha \mathbf{P}(\alpha)/\lambda(q,\alpha)$, where the net transition rate at $\alpha \in \partial(J_{(d)})$ is given in turn by $\lambda(q,\alpha) \equiv \lambda(q,J_{(d)}) := \sum_{j \notin J_{(d)}}q_j,$ i.e., the net transition rate is constant on each ``facet" $\partial(J_{(d)})$ of $X$. Thus
\begin{equation}
\label{eq:tinfq}
t_{\infty} = \frac{1}{z(\eta)}\sum_{d}\sum_{J_{(d)}}\frac{1}{\lambda(q,J_{(d)})}\sum_{\alpha \in \partial(J_{(d)})}\eta^\alpha,
\end{equation}
and the RHS can be evaluated using \eqref{eq:CHSF}.

If $\bm{p}$ denotes the invariant distribution on $X$, then a brief calculation yields $\lVert \bm{p} \rVert = z^{1/2}(\eta^2)/z(\eta)$. Moreover, \eqref{eq:betaEk} yields $\beta E_\alpha = \left \langle \alpha_0 - \alpha, \log \eta \right \rangle$, where $\alpha_0 := z(1)^{-1}\sum_{\alpha}\alpha$ and the logarithm is taken componentwise.

In order to obtain a closed-form expression for $\Theta$, it is necessary to evaluate the sum $\sum_\alpha (\beta E_\alpha)^2$ in closed form. It is easy to see that this reduces to evaluating $\sum_\alpha \langle \alpha, h \rangle^2$ for $h$ a generic $B$-tuple. This can in turn be accomplished given a formula for $\sum_{k=0}^{u}k^w\binom{s+k}{k}$ with $s$, $u$, $w$ integral: and this may finally be obtained by writing $k^w$ as a linear combination of binomial coefficients and considering the sums $\sum_{k=0}^{u}\binom{k}{v}\binom{s+k}{k}$ for $ 0 \leq v \leq w$ \cite{E}. The relevant calculations are performed in an appendix.

So, $\Theta$ can be evaluated in closed form for the closed queue along the lines sketched above. The final result has a lengthy and unenlightening expression, and so we content ourselves here with the simple case of $B = 2$ and $b > 1$. Writing $r := R_{21}/R_{12}$, we have that for $b$ large (so that finite size effects do not dominate) and $q_2 \equiv 1$ (and in fact more generally), $\Theta$ is greatest when the dynamics approaches that of an unbiased random walk: here, this is when $q_1 \approx r$.

\begin{figure}[htbp]
\includegraphics[width=80mm,keepaspectratio]{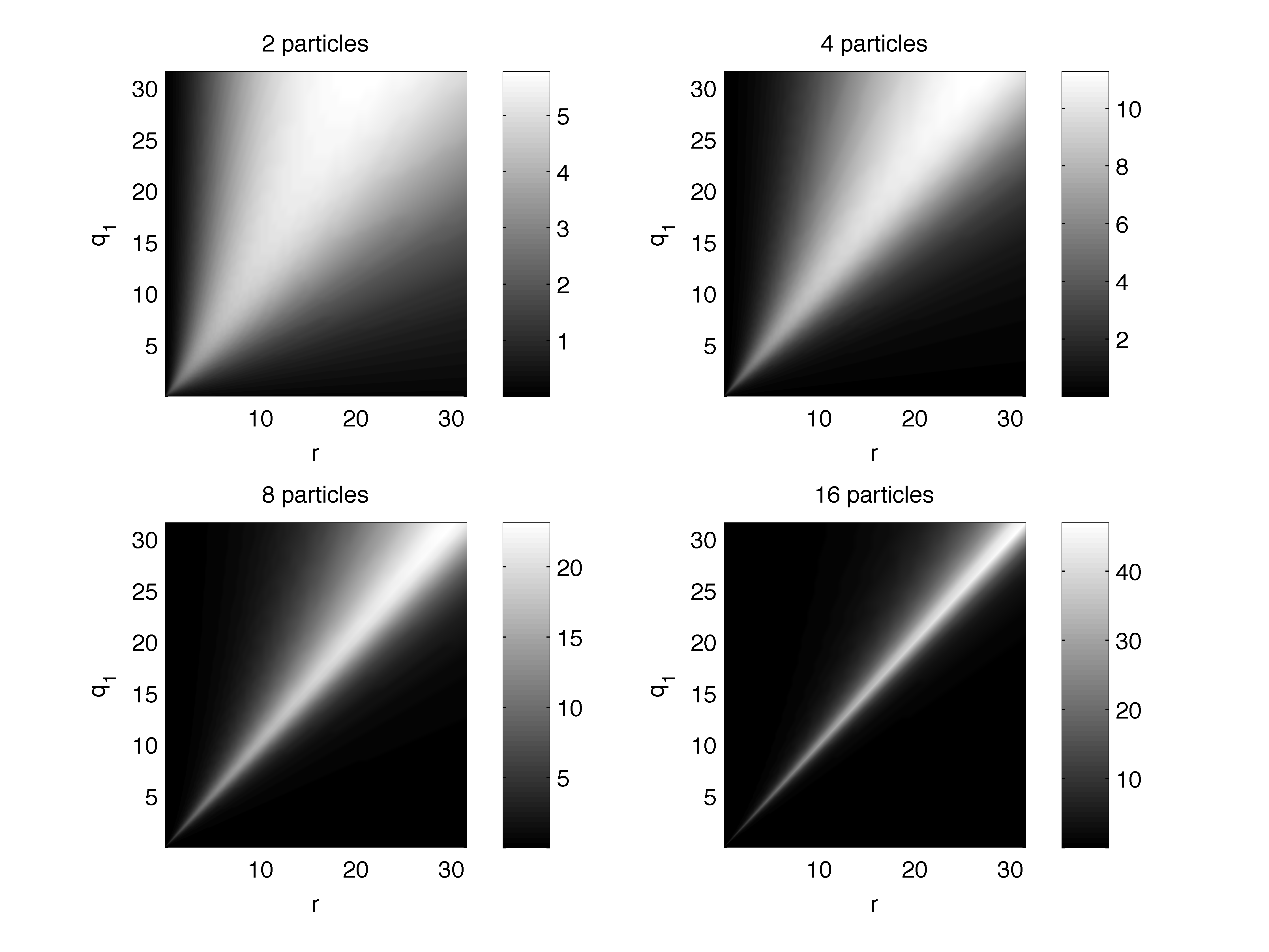}
\caption{   \label{fig:semi} Psuedo-grayscale plot of $\log \Theta$ for $B = 2$ and $b = 2,4,8,16$ as a function of $r$ and $q_1$ with $q_2 \equiv 1$.}
\end{figure}

\paragraph{Discussion.}

We shall make a few general points before proceeding with a further application to computer network traffic monitoring. 

First, if the system under consideration is not stationary, so that $\{p_k\}$ and $t_{\infty}$ vary with time (but are presumably still well-defined), then so will $\Theta$ and $\{E_k\}$, but the language of equilibrium statistical physics is still adequate \emph{even if the system is not actually in equilibrium}. That is, there is no need for (e.g.) detailed balance or a maximum-entropy variational principle to be satisfied in order for $\Theta$ to be well-defined: equation \eqref{eq:temperature} can be taken as an \emph{extension} of the language of equilibrium statistical physics. Though the details of how $\{p_k\}$ and $t_{\infty}$ should be calculated from theory or estimated from measurements are important and nontrivial, these may be neglected here. Once we have $\{p_k\}$ and $t_{\infty}$, we may treat the system at any instant \emph{as if} it were in a stationary state of equilibrium.

Second, many systems of interest in statistical physics are specified by an energy function (in which case there is usually no need for the formalism in this paper), rather than some more detailed rule such as a \emph{bona fide} Hamiltonian system that allows us to compute the dynamical evolution of the system and its characteristic recurrence time. A notable exception is the Ising model with Glauber dynamics, and the application of the formula \eqref{eq:temperature} to a Glauber spin is discussed elsewhere \cite{F2}. 

Third, it is not obvious how to generalize this framework to systems whose state space $\Gamma$ has infinite measure. However if $\Gamma$ is endowed with an appropriate finite measure $\mu$ (as when for example $\Gamma$ is naturally realized as a bounded set in $\mathbb{R}^N$ with smooth boundary, and $\mu$ is induced from Lebesgue measure), the adaptation is comparatively straightforward provided that $t_\infty$ is well defined and $p$ and $\log p$ are in $L^1 \cap L^2$, since $\lVert \beta E \rVert ^2 = \lVert \log p \rVert ^2 - \frac{1}{\mu(\Gamma)}\ \int_{\Gamma} \log p \, d\mu$.

Fourth, the vast majority of finite systems of potential interest are amenable to measurement first, and calculation second, if at all. This raises the prospect of using \eqref{eq:temperature} as a tool for data analysis in the same spirit as entropies are often employed. The potential to apply similar ideas to computer network traffic has been considered by others (see, e.g. \cite{Burgess}) and an application is sketched below.

Finally, in those cases where we can measure everything, it is still possible and generally also preferable to deal with coarsened quantities instead. Then the problem of determining a proper coarsening to work with arises. This is a direct analogue of the Gibbs paradox (where the value of the entropy depends on the level of description of the system). \cite{J2} In applications such as the one considered below, these sorts of issues will typically loom large.

\paragraph{Application to computer network traffic analysis and characterization.}

The Bose gas framework can be applied to traffic analysis provided that each of the $B$ states corresponds to a (type of) source or destination in a communications or other network. Time-inhomogeneous rates $q_j$ and probabilities $R_{jk}$ can be readily estimated from the number of connections with origin (type) $j$ and destination (type) $k$, and this procedure in effect couples the traffic to a Bose gas. \cite{FH} In general, most traffic data sets of interest will have a large number of physical or logical addresses and additional metadata which must be aggregated in order to provide a meaningful context for the traffic and to keep $B$ reasonably small. A simple example of this sort of aggregation for computer network traffic will be sketched below.

Both the Gibbs paradox and the renormalization group inform the aggregation process. By effectively separating both internal levels of description and external scales, the bulk traffic can generally be understood using relatively few parameters, and traffic exhibiting greater complexity can be examined separately and in more detail. The assertion that bulk traffic can be described with a few parameters amounts to saying that the bulk traffic is understandable in the first place, and can be expected to follow from the fact that a communications or other network of interest for traffic analysis is engineered (even if the engineering process is distributed or collaborative) and its general function is knowable at the outset. 

The separation of external scales is essentially the converse of data fusion, where the idea is to identify states at different sensors in an appropriate fashion and compute the corresponding fused rates $q$ and probabilities $R$. In practice this procedure may be easier than it might seem at first. For example, each of several streams of multiplexed computer network traffic may be best examined using the same source/destination types, with identically defined tables of observed ports, network addresses, or communications sockets that are individually populated on a per-stream basis. In this case the states for each traffic stream are identical, and if the $a$th stream has $N_{jk}^{(a)}$ transitions from state $j$ to state $k$ during the time interval $[t, t+\Delta t)$, then setting $N_{jk} := \sum_a N_{jk}^{(a)}$ and (after suitable conditioning if necessary) $q_j := \sum_k N_{jk} / \Delta t$ and $R_{jk} := N_{jk} / \sum_k N_{jk}$ effects a simple data fusion procedure. Note that (all other things being equal) the effective temperature of data fused in this way will scale as the number of fused traffic streams, but this scaling can be offset by dividing $q_j$ by the number of fused streams.

In order to keep the remainder of the discussion nominally self-contained, we provide a brief overview of the types of computer network traffic considered here. For more information on network protocols, see \cite{Stevens}. 

The Internet Protocol (IP) is a connectionless network-layer protocol responsible for routing network packets between sources and destinations. The Internet Control Message Protocol (ICMP) is a network/transport-layer protocol, though ICMP messages are encapsulated within IP datagrams. Comforming ICMP messages can be categorized as either requests, responses, or errors. The User Datagram Protocol (UDP) is a connectionless transport-layer protocol, commonly associated with the Domain Name Service (DNS), Simple Network Management Protocol (SNMP), streaming media (e.g., voice over IP), etc. Although UDP is connectionless, notification of packet reception errors over ICMP is a common protocol behavior (though its enabling or disabling depends on local network policy). Therefore it is often appropriate to analyze UDP and ICMP in the same framework. Towards that end, and after analyzing benign UDP and ICMP traffic from an independently operated gigabit network testbed simulating a large network enclave connected to the internet, an exhaustive set of states that reflected the bulk features of the network traffic source/destination pair attributes was formulated. 

Each UDP or ICMP packet, along with its encapsulating IP header, corresponds to an attempted state transition of a single-particle Bose gas. (Multi-particle Bose gases may also be considered; however, increasing the particle number can increase both artificial ``crosstalk" [which may or may not be desirable] and noise.) IP addresses along with UDP ports or ICMP types corresponding to well-known services, ephemeral client-side connections, or message types were considered as source and destination attributes defining the internal states according to the figure below. The Bose gas statistics were autonomously updated approximately every second (a ``stopping time" update protocol that bounds update intervals from below was actually used).

\begin{figure}[htbp]
\includegraphics[width=80mm,keepaspectratio]{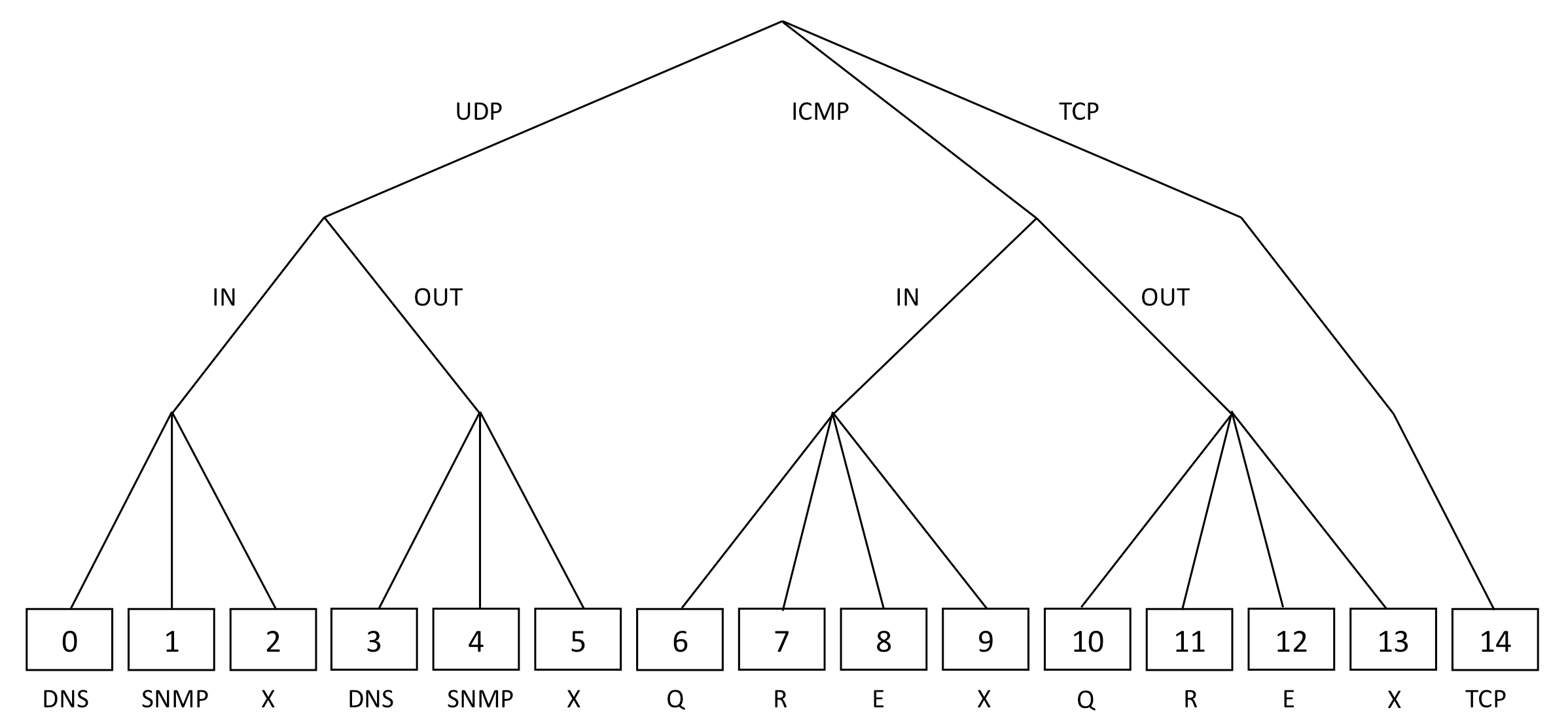}
\caption{   \label{fig:semi} Packet classification tree for source and destination types. ``X" is a shorthand for ``other"; ``Q", ``R", and ``E" for ICMP requests, responses, and errors, respectively. Although not done for the data presented here, it is convenient for a number of reasons to force ICMP with source attributes involving X to have destination attributes involving E and vice versa, and similarly for Q and R. Also note that the set of states is automatically exhaustive by virtue of the tree structure.}
\end{figure}

Sophisticated classes of internal states can be defined and implemented, e.g., ``IP address inside the network and present in observed traffic between 1 and 10 times during the last approximately 5 seconds; UDP port not present in observed traffic during the last approximately 1 minute", but the dynamic nature of such states effectively precludes the sort of analysis presented here, which relies on memoryless source/destination types. 

The bulk of benign UDP and ICMP traffic was DNS, and the well-known Slammer worm was introduced onto the testbed during the time interval indicated by the shaded areas on the following figures. Although the traffic levels associated with the worm were found to be an order of magnitude greater than the benign traffic, the concomitant source and destination state pairs were distinct from the benign traffic in such a way as to enable the determination of temperature (as well as entropy) characteristics based on the presence of all, a portion, or none of the worm traffic. (Furthermore, a slight but detectable change in the effective temperature slightly preceded the corresponding change in traffic rate.) This in turn allowed us to determine that the effective temperature was significantly more robust as a discriminator of the worm activity than the entropy, as demonstrated by the simple expedient of removing the ICMP errors caused by the worm (and responsible for roughly half of the offending traffic) as well as varying amounts of the actual worm traffic itself via post-processing.

\begin{figure}[htbp]
\includegraphics[width=80mm,keepaspectratio]{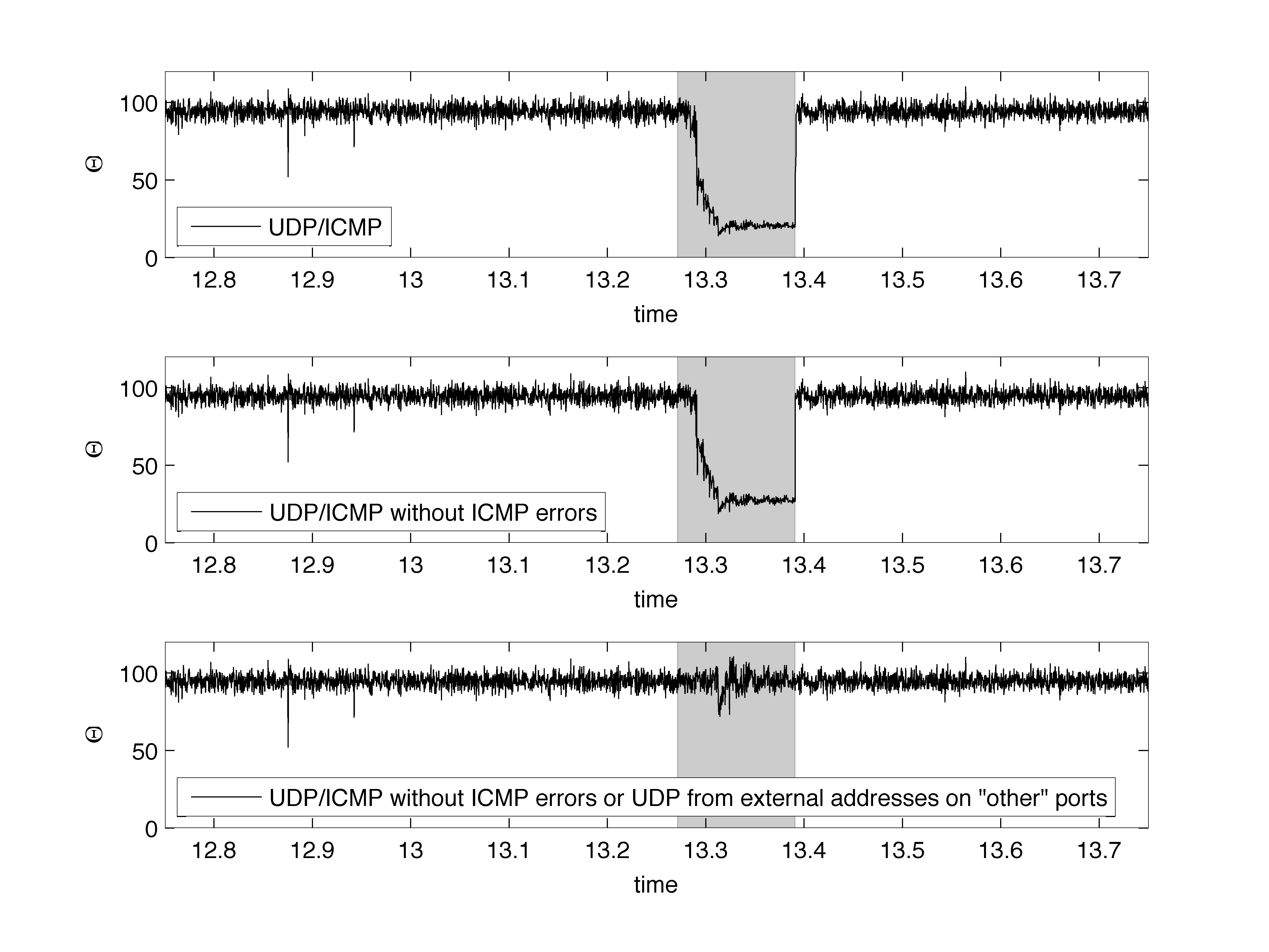}
\caption{   \label{fig:semi} Effective temperature of a one-particle Bose gas coupled to a network testbed. The period of Slammer worm activity is highlighted. Time in this and similar figures is in hours.}
\end{figure}

\begin{figure}[htbp]
\includegraphics[width=80mm,keepaspectratio]{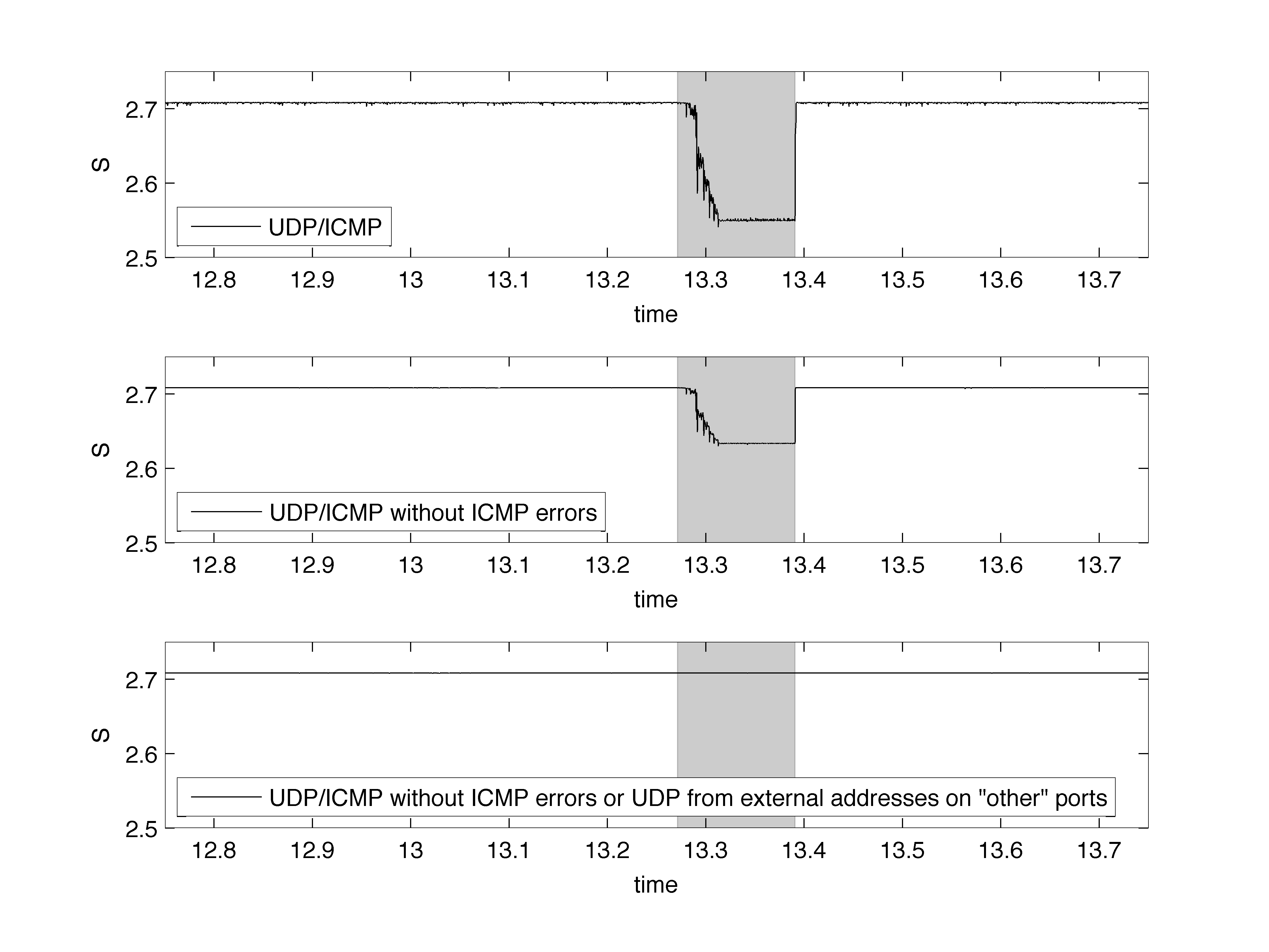}
\caption{   \label{fig:semi} Entropy of a one-particle Bose gas coupled to a network testbed. The period of worm activity is highlighted.}
\end{figure}

\begin{figure}[htbp]
\includegraphics[width=80mm,keepaspectratio]{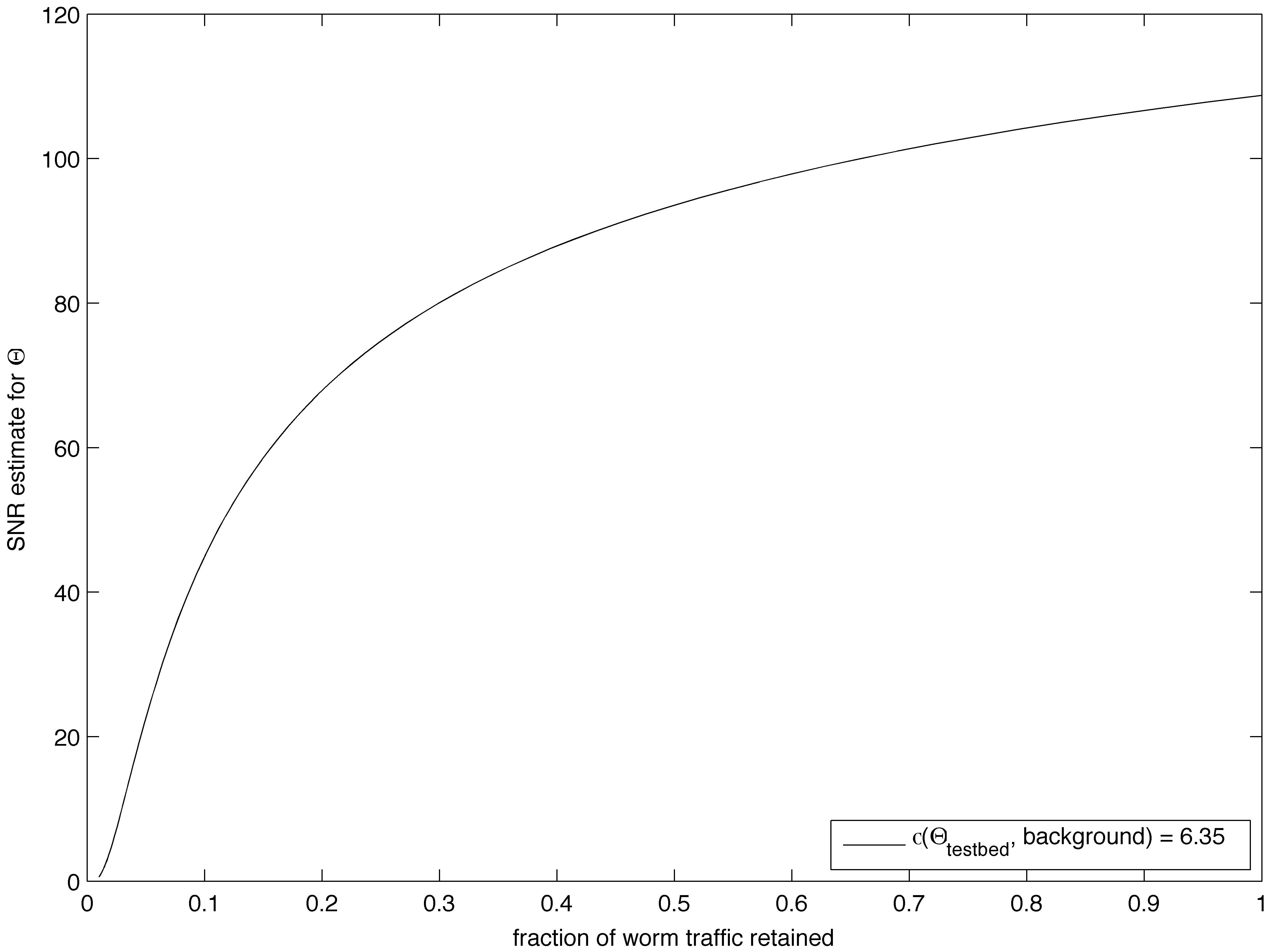}
\caption{   \label{fig:semi} Signal to noise ratio behavior of the effective temperature.}
\end{figure}

At a finer level of detail, anomalous network activity was detectable through analysis of certain of the power components
\begin{equation}
\dot W_j = \left ( \frac{b}{B} - \langle \alpha \rangle_j \right ) \cdot D_t (\Theta \log \eta_j)
\end{equation}
where $\sum_j \dot W_j \equiv \dot W \equiv \sum_\alpha p_\alpha \dot E_\alpha$ and
\begin{equation}
\dot U = \sum_\alpha \dot p_\alpha E_\alpha + \sum_\alpha p_\alpha \dot E_\alpha \equiv \dot Q + \dot W.
\end{equation}

In particular, we show two of the integrated components for testbed UDP/ICMP traffic with all the periods of malicious activity on the testbed highlighted. The worm corresponds to the penultimate period, a set of scans to the period before that, and the remainder of periods to other types of malicious traffic, generally involving a handful of packets. The components shown below are significantly more effective discriminators of anomalous activity than most others.

\begin{figure}[htbp]
\includegraphics[width=80mm,keepaspectratio]{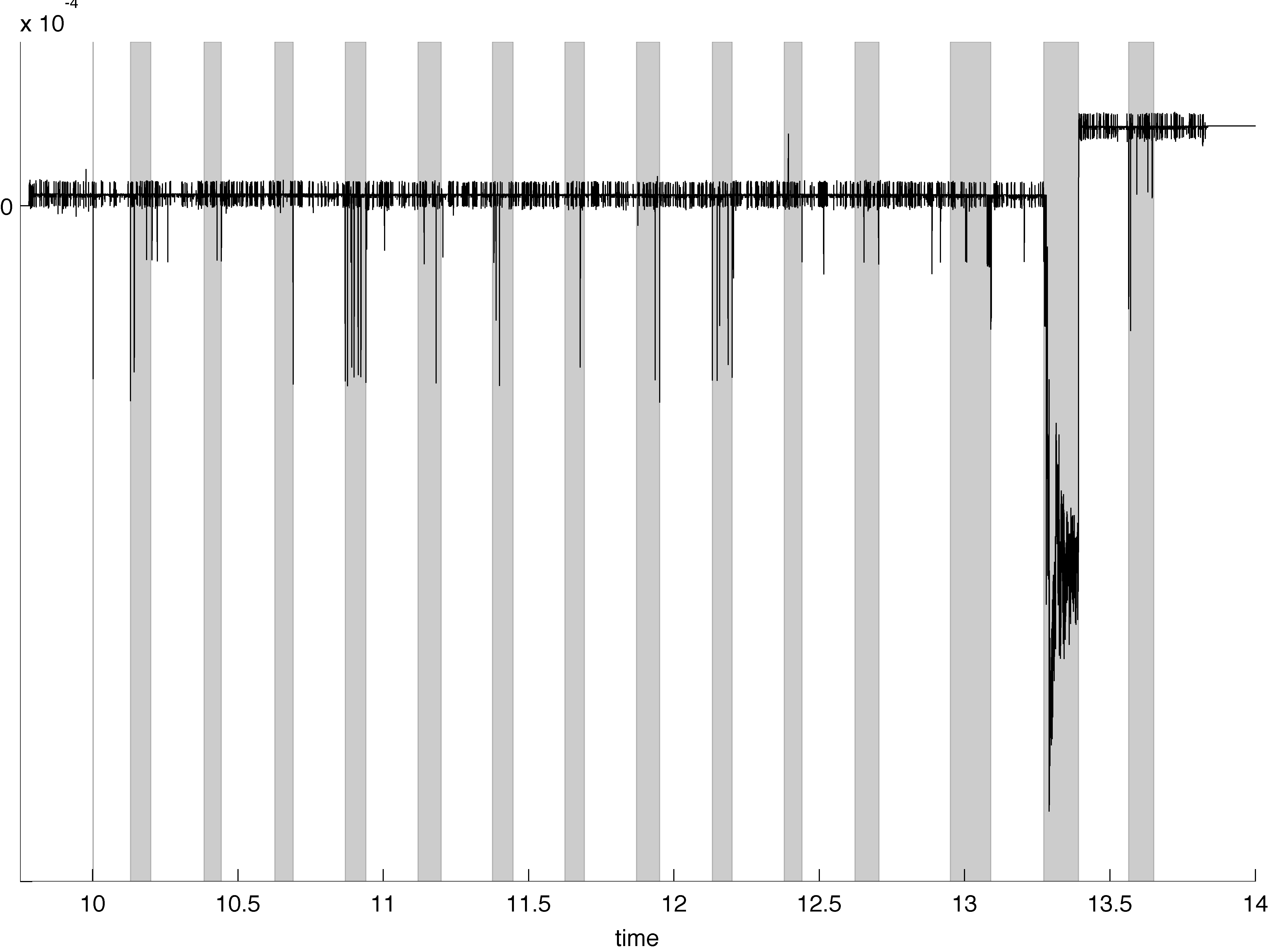}
\caption{   \label{fig:semi} A work component of the one-particle UDP Bose gas. Each of the larger spikes in this or similar figures either occurred simultaneously with a malicious packet or at a time for which log file data from the testbed was later found to be unavailable.}
\end{figure}

\begin{figure}[htbp]
\includegraphics[width=80mm,keepaspectratio]{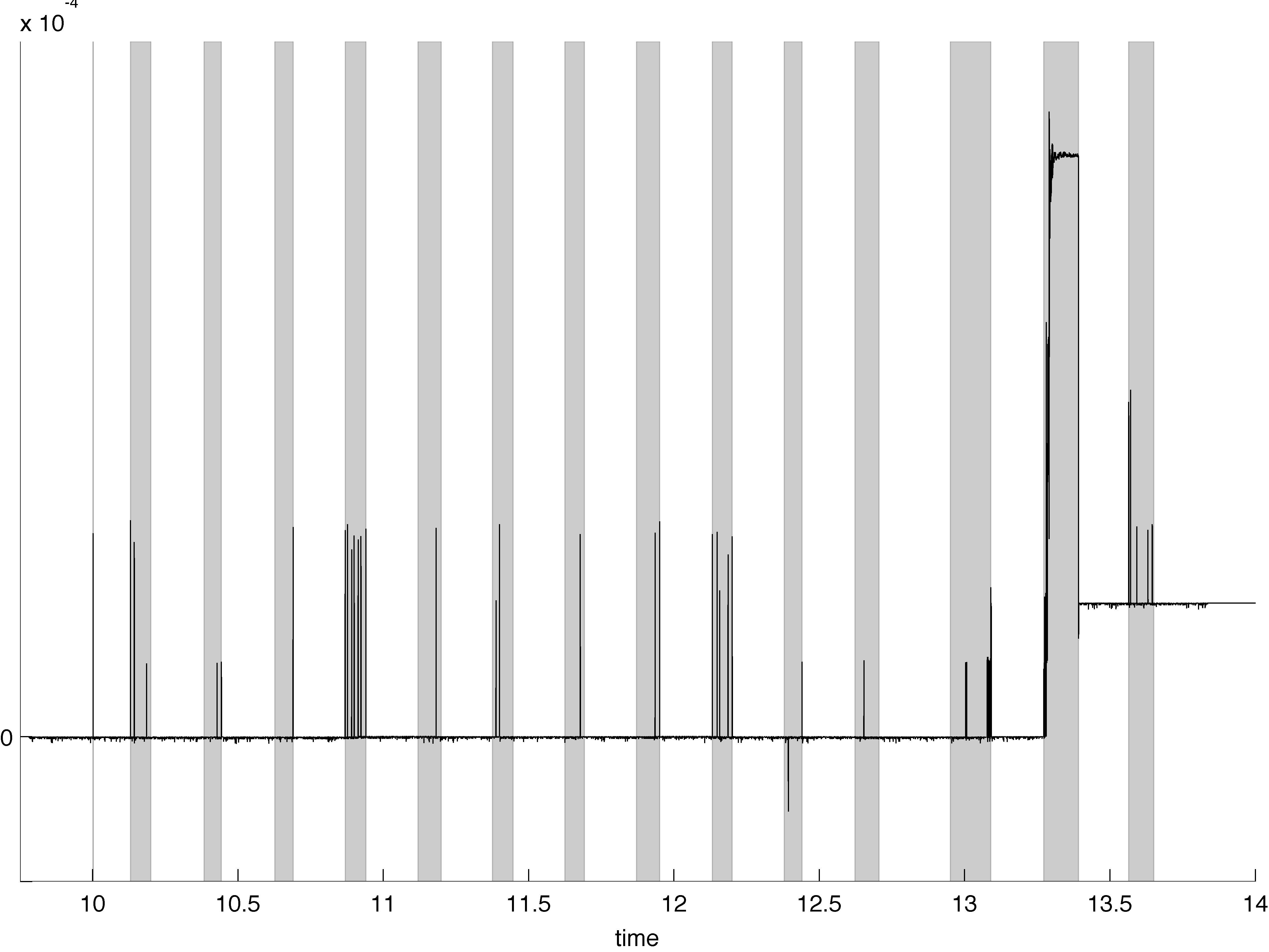}
\caption{   \label{fig:semi} Another work component of the one-particle UDP Bose gas.}
\end{figure}

Next, we show two of the more effective integrated components for testbed Transmission Control Protocol (TCP) traffic, with a similar classification scheme in which mail, web, FTP, and two other service ports were differentiated; all other low ports were aggregated, as were all high ports. To accommodate the static nature of these decision tree nodes for analytical purposes, FTP sessions from high ports to high ports were disabled, though a tree node taking recent IP addresses into account would more than compensate for any effects of this.

\begin{figure}[htbp]
\includegraphics[width=80mm,keepaspectratio]{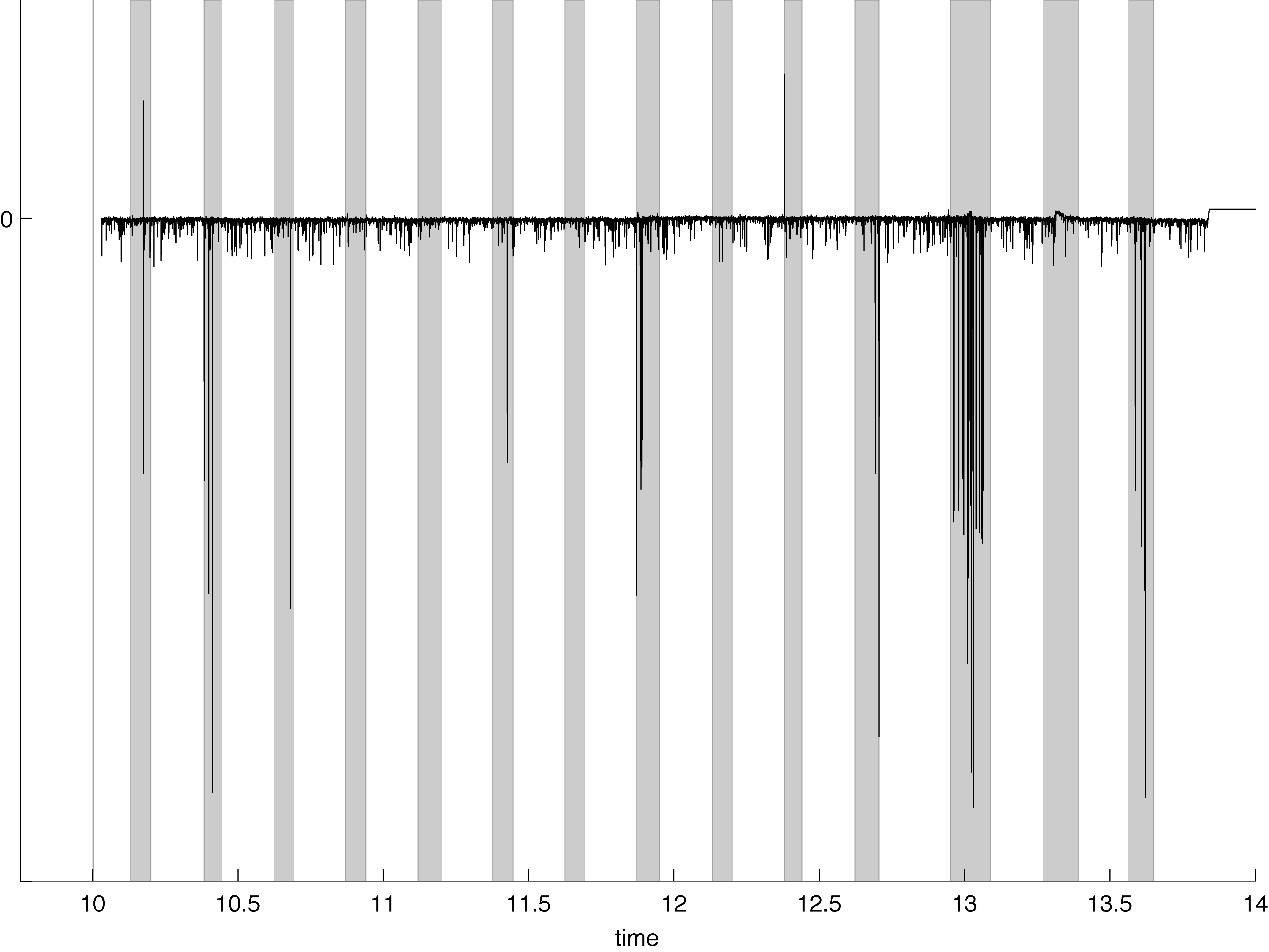}
\caption{   \label{fig:semi} A work component of the one-particle TCP Bose gas. The lack of activity during some periods of malicious traffic besides the worm interval can be attributed to the simple memoryless nature of the classification procedure used here.}
\end{figure}

\begin{figure}[htbp]
\includegraphics[width=80mm,keepaspectratio]{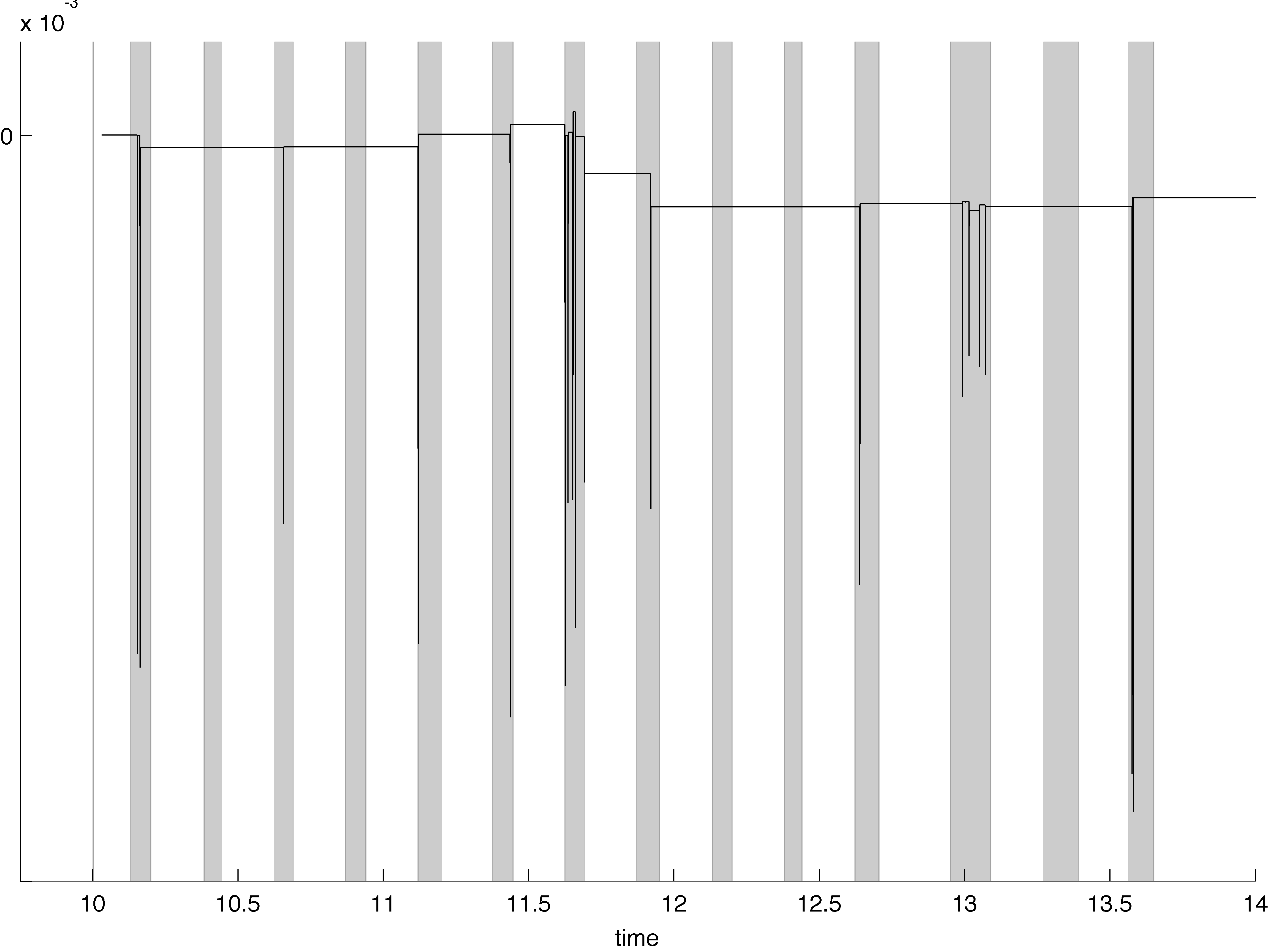}
\caption{   \label{fig:semi} Another work component of the one-particle TCP Bose gas.}
\end{figure}

In particular, the lack of accounting for connection or flow status (TCP is a connection-oriented protocol), address and/or port frequencies, the number of bytes transferred, TCP sequence and acknowledgement numbers, etc. means that malicious traffic directed at internal web or email servers from outside the enclave would generally be difficult to distinguish from genuinely anomalous or malicious traffic directed to internal addresses on web or email ports, but that is only due to the simple nature of the classification tree in this example. In practice it would generally be advantageous to separate web and email traffic into streams for detailed analysis (possibly including states tied to the application layer of the protocol stack), or at least to treat web and email servers differently in the source/destination type association scheme. 

Although such a procedure was not carried out here (partially in order to keep the analysis presented here comparatively straightforward), note that the UDP/ICMP and TCP work components here combine to unambiguously indicate malicious activity in each of the highlighted periods through a simple thresholding procedure. During those periods where sufficiently accurate logs detailing the malicious traffic were available (the entire period for UDP/ICMP and the last period of malicious traffic for TCP) each of the values of these work components above appropriate thresholds occurred at a timestamp for malicious activity. While this sort of approach does not in itself determine the difference between malicious and anomalous traffic, an interactive visual traffic analysis engine and autonomous offloaded domain-specific methods can both aid the characterization of normal traffic and perform more contextual analysis of any anomalous traffic. \cite{Huntsmanprep} This suggests the possibility of a practical technique for network traffic monitoring that incorporates interactive configuration (i.e., refinements of classification trees combined with separation of traffic streams and timescales, all mediated through queries of, e.g., the routing time series $R_{jk}$ down to the level of individual transactions) followed by automated detection with low computational requirements and attractive performance characteristics.

\paragraph{Conclusion.}

Our discussion turns at last to a few concluding remarks. Though the usual state of affairs is for temperature to be regarded as an environmental parameter in calculations, the logic can be turned on its head: through the simple expedient of measuring occupation times, we can directly obtain an effective temperature (and the corresponding energy levels) of a finite system without recourse to a Hamiltonian, and typically using appropriate observables of our choosing. Without a Hamiltonian, we cannot predict the system's exact microscopic evolution, but we can still efficiently describe the macroscopic evolution using the idiom of equilibrium statistical physics. It is desirable to describe systems in this way precisely because this approach has proven so successful at ignoring irrelevant microscopic details while still providing the most relevant information about a system. While a proper interpretation of an effective temperature in purely information-theoretical terms is not yet known (and a recent philosophical study of thermometry notes that ``[in] fact, there are complicated philosophical disputes about just what kind of quantity temperature is" \cite{Chang}), we believe that the connection between information theory and statistical physics may eventually be extended to encompass temperature and energy as well. 

\appendix
\section{Evaluating a certain sum}

Let $h \in \mathbb{R}^B$ and consider the sum
\begin{equation}
\label{eq:sumah2}
\sum_{\alpha} \langle \alpha , h \rangle^2.
\end{equation}
We have that 
\begin{equation}
\sum_{\alpha} \langle \alpha , h \rangle^2 = \sum_j h_j^2 \sum_\alpha \alpha_j^2 + \sum_{j\ne k} h_j h_k \sum_\alpha \alpha_j \alpha_k,
\end{equation}
so in order to evaluate the LHS it suffices to compute
\begin{equation}
\sigma_2 := \sum_\alpha \alpha_j^2
\end{equation}
and
\begin{equation}
\sigma_{11} := \sum_\alpha \alpha_j \alpha_k \quad (j \ne k)
\end{equation}
(note that there is no real dependence on $j$ or $k$ in the RHS of either of the preceding two equations).

The $\sigma$-coefficients can be readily evaluated by first evaluating sums of the form
\begin{equation}
\sum_{k=0}^n k^p \binom{r+k}{k}
\end{equation}
with $p$ a positive integer. This in turn is accomplished by writing $k^p$ as a linear combination of binomial coefficients and considering 
\begin{equation}
\sum_{k=0}^n \binom{k}{q} \binom{r+k}{k}
\end{equation}
for $0 \le q \le p$ \cite{E}. The relevant results of this procedure for the present purpose are
\begin{equation}
\label{eq:binom0}
\sum_{k=0}^n \binom{r+k}{k} = \binom{n+r+1}{r+1}
\end{equation}
\begin{equation}
\label{eq:binom1}
\sum_{k=0}^n k\binom{r+k}{k} = (r+1) \cdot \binom{n+r+1}{r+2}
\end{equation}
and
\begin{equation}
\label{eq:binom2}
\sum_{k=0}^n k^2 \binom{r+k}{k} = ((r+2)n+1) \cdot \frac{r+1}{r+3} \cdot \binom{n+r+1}{r+2}.
\end{equation}

Some wholly elementary but tedious manipulations using equations \eqref{eq:binom0}-\eqref{eq:binom2} along with the fact that $z(1) = \sum_\alpha 1 = \binom{B+b-1}{b}$ quickly lead to the results
\begin{equation}
\label{eq:sigma2}
\sigma_2 = \frac{B+2b-1}{b-1} \sigma_{11}
\end{equation}
where
\begin{equation}
\label{eq:sigma11}
\sigma_{11} = \binom{B+b-1}{B+1}.
\end{equation}
We will briefly outline these manipulations while omitting a few trivial steps for the case of equation \eqref{eq:sigma2}. 

Now
\begin{equation}
\sigma_2 = \sum_\alpha \alpha_1^2 = \sum_{\alpha_1 = 0}^b \alpha_1^2 z_{B-1,b-\alpha_1}(1), 
\end{equation}
so
\begin{equation}
\sigma_2 = \sum_{s=0}^b s^2 \binom{B-2+b-s}{b-s} = \sum_{l=0}^b (b-m)^2 \binom{r+m}{m}
\end{equation}
where we have made the substitutions $r := B-2, m:= b-s$. Expanding the squared term and using the forumlae above before undoing the substitutions yields
\begin{equation}
\begin{aligned}
\sigma_2 = b^2 \binom{B+b-1}{B-1} + (1-b(B+2))\frac{B-1}{B+1} \binom{B+b-1}{B}.
\nonumber
\end{aligned}
\end{equation}
Finally, a few lines of algebra yields the result \eqref{eq:sigma2}. Equation \eqref{eq:sigma11} follows similarly.

Summarizing, we have that \eqref{eq:sumah2} equals
\begin{equation}
\binom{B+b-1}{B+1} \cdot \left ( \sum_{j\ne k} h_j h_k + \frac{B+2b-1}{b-1} \sum_j h_j^2 \right ).
\end{equation}

\section{Invariant distribution of the classical Bose gas}

If $Q$ is the generator of a well-behaved continuous-time Markov process on a finite state space, its invariant distribution $p$ is the unique solution of the eigenproblem $pQ = 0$ satisfying $\left| p \right| = 1$.

In the case of the classical Bose gas, the nontrivial entries of the generator are given by \eqref{eq:generator}. Define
\begin{equation}
\lozenge(\alpha) := \{(j,k) : j \ne k \land \alpha - e_j + e_k \in X_{B,b}\}
\end{equation}
and note that $\lozenge(\alpha) = \{(j,k) : j \ne k \land \alpha_j \ne 0\}$. Since the rows of the generator must sum to zero, we have
\begin{equation}
-Q_{\alpha,\alpha} = \sum_{(j,k) \in \lozenge(\alpha)} Q_{\alpha,\alpha - e_j + e_k} = \sum_{(j,k) \in \lozenge(\alpha)} q_j R_{jk}.
\end{equation}

Now $(pQ)_\alpha = 0$ iff
\begin{equation}
-p_\alpha Q_{\alpha,\alpha} = \sum_{(j,k) \in \lozenge(\alpha)} p_{\alpha - e_j + e_k} Q_{\alpha - e_j + e_k,\alpha}
\end{equation}
or equivalently
\begin{equation}
p_\alpha \sum_{(j,k) \in \lozenge(\alpha)} q_j R_{jk} = \sum_{(j,k) \in \lozenge(\alpha)} p_{\alpha - e_j + e_k} q_k R_{kj}.
\end{equation}
Under the \emph{Ansatz} $p_\alpha = \eta^\alpha/z$, this takes the form
\begin{equation}
\sum_{(j,k) \in \lozenge(\alpha)} q_j R_{jk} = \sum_{(j,k) \in \lozenge(\alpha)} \frac{\eta_k}{\eta_j} q_k R_{kj}.
\end{equation}
Writing the summation indices explicitly and recalling that $\eta = \pi/q$, we obtain
\begin{equation}
\sum_{j:\alpha_j \ne 0} q_j \sum_{k \ne j} R_{jk} = \sum_{j:\alpha_j \ne 0} \frac{1}{\eta_j} \sum_{k \ne j} \pi_k R_{kj}.
\end{equation}
Using $\pi R = \pi$ and $\sum_{k \ne j} R_{jk} = 1 - R_{jj}$ this simplifies to
\begin{equation}
\sum_{j:\alpha_j \ne 0} q_j (1 - R_{jj}) = \sum_{j:\alpha_j \ne 0} \frac{1}{\eta_j} \pi_j (1 - R_{jj}).
\end{equation}
At this point equality is clear, establishing that $p_\alpha = \eta^\alpha/z$ is in fact the invariant distribution.

\begin{acknowledgments}
The author thanks Ronald Fisch, David Ford, John Franklin, James Heagy, Brian Hearing, and James Luscombe for their helpful guidance and corrections; the Johns Hopkins University Applied Physics Laboratory for access to their network testbed, the Hewlett-Packard Company for preprocessing the data presented here, and finally to Christopher Covington and Greg Tumolo for processing and formatting the data. This work was partially supported by DARPA.
\end{acknowledgments}


\bibliography{Temperaturepaperbib}

\end{document}